\documentclass[aps,rpl,preprint,groupedaddress]{revtex4}
\begin{document}

\title{Quantum channels with correlated noise and entanglement teleportation}
\author{Ye Yeo}

\affiliation{Centre for Mathematical Sciences, Wilberforce Road, Cambridge CB3 0WB, United Kingdom}

\begin{abstract}
Motivated by the results of C. Macchiavello and G. M. Palma on entanglement-enhanced information transmission over a quantum channel with correlated noise, we demonstrate how the entanglement teleportation scheme of J. Lee and M. S. Kim gives rise to two uncorrelated generalized depolarizing channels.  In an attempt to find a teleportation scheme which yields two correlated generalized depolarizing channels, we discover a novel teleportation scheme, which allows one to learn about the entanglement in an entangled pure input state, but without decreasing the amount of entanglement associated with it.
\end{abstract}

\maketitle

In quantum mechanics, the formalism of quantum operations, described in detail by Kraus \cite{Kraus}, describes the most general possible state change.  In this formalism there is an input state $\rho_{in}$ and an output state $\rho_{out}$, which are connected by a map
$$
\rho_{in} \longrightarrow \rho_{out} 
= \frac{{\cal E}(\rho_{in})}{{\rm tr}[{\cal E}(\rho_{in})]}.
$$
The map is determined by a quantum operation $\cal E$, a linear, trace-decreasing map that preserves complete positivity.  The most general form for $\cal E$ can be shown to be \cite{Kraus}
$$
{\cal E}(\rho_{in}) = \sum_k A_k\rho_{in}A^{\dagger}_k,
$$
\begin{equation}
\sum_k A^{\dagger}_kA_k \leq I.
\end{equation}
The Kraus operators $A_k$ completely specify the quantum operation $\cal E$.  The insight \cite{Nielsen, Caves} that one can recast the problem of achieving optimal quantum teleportation \cite{Bennett} into one of optimal reversal of quantum operations, enables one to establish a more general connection between quantum operations, or quantum channels, and teleportation schemes.  In Ref.\cite{Yeo}, it was demonstrated how this connection enables one to explore optimal approximate reversal of quantum operations on a single qubit.  

Recently, the problem of classical capacity of quantum channels with time correlated noise was considered by C. Macchiavello and G. M. Palma \cite{Mac}.  In particular, the problem of subjecting quantum states to two correlated or uncorrelated depolarizing channels were analyzed.  In the light of the above insight, it is natural to ask what teleportation schemes would give rise to two correlated or uncorrelated generalized depolarizing channels.  In this paper, we show how the entanglement teleportation scheme of J. Lee and M. S. Kim \cite{Lee} yields two uncorrelated generalized depolarizing channels.  However, in order to obtain two correlated generalized depolarizing channels, the entangled states shared between the sender and receiver have to be changed.  Interestingly, this modification yields an entanglement teleportation scheme, which allows one to learn about the entanglement of a bipartite pure input state, decreasing the fidelity \cite{Jozsa} of the teleported state with respect to the input state, but remarkably without changing the amount of entanglement associated with the original input state.

Here, we adopt the following measure of entanglement \cite{Lee}.  Consider a density operator $\tilde{\rho}_{12}$ and its partial transposition $\tilde{\sigma}_{12} = \tilde{\rho}^{T_1}_{12}$ for two 2-level quantum systems.  The density operator $\tilde{\rho}_{12}$ is entangled if and only if $\tilde{\sigma}_{12}$ has any negative eigenvalues.  The measure of entanglement $E(\tilde{\rho}_{12})$ defined by
$$
E(\tilde{\rho}_{12}) = \max(0, -2\sum_a\lambda^-_a),
$$
where $\lambda^-_a$ is a negative eigenvalue of $\tilde{\sigma}_{12}$, satisfies the necessary conditions required for every measure of entanglement.

To set the stage, we begin with a description of a general teleportation scheme, which involves a sender, Alice, and a receiver, Bob, sharing a single pair of entangled particles.  Alice is in possession of two $n$-level quantum systems, the input system $1$, and another system $2$ arbitrarily entangled with a third $n$-level target system $3$ in Bob's possession.  Initially the composite system $123$ is prepared in a state with density operator $\tilde{\rho}_1 \otimes \chi_{23}$, where $\tilde{\rho}_1$ is an unknown state of the input system $1$, and $\chi_{23}$ is an arbitrary entangled state of systems $2$ and $3$.  Since the systems $1$ and $3$ are identical and thus have the same state space, a one-to-one correspondence from the state space of the composite system onto itself can be established by a unitary swap operator $U_{1(2)3}$, which acts on product states according to
$$
U_{1(2)3}(|\tilde{a}\rangle_1 \otimes |b\rangle_2 \otimes |c\rangle_3) =
|\tilde{c}\rangle_1 \otimes |b\rangle_2 \otimes |a\rangle_3,
$$
swapping the states of systems $1$ and $3$, while leaving system $2$ alone.  $U_{1(2)3}$ obviously satisfies $(U_{1(2)3})^2 = I_{123}$, the identity operator on the composite system, and $U^{\dagger}_{1(2)3} = U_{1(2)3}$.  When extended to operators $Q_{123}$ on the composite system, the correspondence becomes
$$
\tilde{Q}_{123} \leftrightarrow Q_{123} 
= U_{1(2)3}\tilde{Q}_{123}U^{\dagger}_{1(2)3}.
$$
It follows that
\begin{equation}
\tilde{\rho}_1 \otimes \chi_{23} = U_{1(2)3}(\tilde{\chi}_{12} \otimes \rho_3)
U^{\dagger}_{1(2)3},
\end{equation}
where $\tilde{\chi}_{12}$ is the counterpart of $\chi_{23}$.

To teleport the input state $\tilde{\rho}_1$ to Bob's target system $3$, Alice performs a generalized measurement on systems $1$ and $2$.  This generalized measurement is described by operators $\tilde{\Pi}^{ij}_{12} \otimes I_3$, where $\tilde{\Pi}^{ij}_{12}$ are Kraus operators on the joint system $12$, $i$ labels the outcome of the measurement, and
$$
\sum_i\sum_j \tilde{\Pi}^{ij\dagger}_{12}\tilde{\Pi}^{ij}_{12} = I_{12}.
$$
If Alice's measurement has outcome $i$, she communicates her measurement result to Bob via a classical channel.  The state of Bob's target system $3$ conditioned on Alice's measurement result $i$ is given by
\begin{equation}
\rho^i_3 = \frac{1}{p_i}{\rm tr}_{12}\left[
\sum_j (\tilde{\Pi}^{ij}_{12} \otimes I_3)(\tilde{\rho}_1 \otimes \chi_{23})
(\tilde{\Pi}^{ij\dagger}_{12} \otimes I_3)
\right],
\end{equation}
where
$$
p_i = {\rm tr}_{123}\left[
\sum_j (\tilde{\Pi}^{ij}_{12} \otimes I_3)(\tilde{\rho}_1 \otimes \chi_{23})
(\tilde{\Pi}^{ij\dagger}_{12} \otimes I_3)
\right].
$$
Substituting Eq.(2) into Eq.(3) gives
\begin{equation}
\rho^i_3 = \frac{1}{p_i}{\rm tr}_{12}\left[
\sum_j (\tilde{\Pi}^{ij}_{12} \otimes I_3)U_{1(2)3}
(\tilde{\chi}_{12} \otimes \rho_3)
U^{\dagger}_{1(2)3}(\tilde{\Pi}^{ij\dagger}_{12} \otimes I_3)
\right].
\end{equation}
Writing
$$
\tilde{\chi}_{12} = \sum_k q_k|\tilde{s}_k\rangle_{12}\langle\tilde{s}_k|,
$$
where the vectors $|\tilde{s}_k\rangle_{12}$ make up the complete orthonormal set of eigenvectors of $\tilde{\chi}_{12}$ in the joint space of $1$ and $2$, and performing the partial trace of Eq.(4) in the complete orthonormal basis $|\tilde{P}_l\rangle_{12}$ for the joint system $12$ give
$$
\rho^i_3 = \frac{1}{p_i}\sum_{j, k, l}q_k\ 
{_{12}}\langle\tilde{P}_l|
(\tilde{\Pi}^{ij}_{12} \otimes I_3)U_{1(2)3}
(|\tilde{s}_k\rangle_{12}\langle\tilde{s}_k| \otimes \rho_3)
U^{\dagger}_{1(2)3}(\tilde{\Pi}^{ij\dagger}_{12} \otimes I_3)
|\tilde{P}_l\rangle_{12}
$$
$$
= \sum_{j, k, l}\left[\sqrt{\frac{q_k}{p_i}}
{_{12}}\langle\tilde{P}_l|(\tilde{\Pi}^{ij}_{12} \otimes I_3)U_{1(2)3}
|\tilde{s}_k\rangle_{12}\right]\rho_3\left[
\sqrt{\frac{q_k}{p_i}}
{_{12}}\langle\tilde{s}_k|U^{\dagger}_{1(2)3}
(\tilde{\Pi}^{ij\dagger}_{12} \otimes I_3)
|\tilde{P}_l\rangle_{12}
\right].
$$
Therefore, $\rho^i_3$ is related to $\rho_3$ by a quantum operation ${\cal E}^i$:
\begin{equation}
\rho^i_3 = {\cal E}^i(\rho_3) = \sum_m A^{im}_3\rho_3A^{im\dagger}_3,
\end{equation}
where
$$
A^{im}_3 \equiv \sqrt{\frac{q_k}{p_i}}
{_{12}}\langle\tilde{P}_l|(\tilde{\Pi}^{ij}_{12} \otimes I_3)U_{1(2)3}
|\tilde{s}_k\rangle_{12},
$$
\begin{equation}
\sum_m A^{im\dagger}_3A^{im}_3 \leq I_3,
\end{equation}
and the single index $m$ denotes the triple $(j, k, l)$.

Next, we consider the entanglement teleportation scheme of J. Lee and M. S. Kim \cite{Lee}, in a slightly more general setting.  Alice is in possession of four $n$-level quantum systems, the two entangled input systems $1$ and $2$, and another two systems $3$ and $5$, each arbitrarily entangled respectively with a fifth and a sixth $n$-level target systems $4$ and $6$ in Bob's possession.  Initially the composite system $123456$ is prepared in a state with density operator $\tilde{\rho}_{12} \otimes \chi_{34} \otimes \chi_{56}$, where $\tilde{\rho}_{12}$ is an unknown entangled state of the input system $12$, and $\chi_{34}$ and $\chi_{56}$ are arbitrary entangled states of systems $34$ and $56$.  To teleport the input state $\tilde{\rho}_{12}$ to Bob's target systems $46$, Alice performs generalized measurements on systems $13$ and $25$.  These generalized measurements are described by operators $(\tilde{\Pi}^{ij}_{13} \otimes I_4) \otimes (\tilde{\Pi}^{i'j'}_{25} \otimes I_6)$.  If Alice's measurements have outcomes $(i,\ i')$, she communicates her measurement results to Bob via classical channels.  The state of Bob's target systems $46$ conditioned on Alice's measurement results $(i,\ i')$ is formally given by
$$
\rho^{ii'}_{46} = \frac{1}{p_{ii'}}{\rm tr}_{1325}\left[
\sum_{j, j'}(
(\tilde{\Pi}^{ij}_{13} \otimes I_4) \otimes 
(\tilde{\Pi}^{i'j'}_{25} \otimes I_6))
(\tilde{\rho}_{12} \otimes \chi_{34} \otimes \chi_{56})(
(\tilde{\Pi}^{ij}_{13} \otimes I_4) \otimes 
(\tilde{\Pi}^{i'j'}_{25} \otimes I_6))^{\dagger}
\right]
$$
$$
= \frac{1}{p_{ii'}}{\rm tr}_{1325}\left[
\sum_{j, j'}
((\tilde{\Pi}^{ij}_{13} \otimes \tilde{\Pi}^{i'j'}_{25}) \otimes 
(I_4 \otimes I_6))U_{(1)23\hat{4}\hat{5}(6)}
(\tilde{\rho}_{12} \otimes \chi_{34} \otimes \chi_{56})\right.
$$
$$
\left.
U^{\dagger}_{(1)23\hat{4}\hat{5}(6)}
((\tilde{\Pi}^{ij}_{13} \otimes \tilde{\Pi}^{i'j'}_{25}) \otimes 
(I_4 \otimes I_6))^{\dagger}
\right]
$$
$$
= \frac{1}{p_{ii'}}{\rm tr}_{1325}\left[
\sum_{j, j'}
((\tilde{\Pi}^{ij}_{13} \otimes \tilde{\Pi}^{i'j'}_{25}) \otimes 
(I_4 \otimes I_6))U_{1(2)\hat{3}(4)5\hat{6}}
(\tilde{\chi}_{13} \otimes \tilde{\chi}_{25} \otimes \rho_{46})\right.
$$
\begin{equation}
\left.
U^{\dagger}_{1(2)\hat{3}(4)5\hat{6}}
((\tilde{\Pi}^{ij}_{13} \otimes \tilde{\Pi}^{i'j'}_{25}) \otimes 
(I_4 \otimes I_6))^{\dagger}
\right]
\end{equation}
where
$$
p_{ii'} = {\rm tr}_{123456}
\left[
\sum_{j, j'}
((\tilde{\Pi}^{ij}_{13} \otimes \tilde{\Pi}^{i'j'}_{25}) \otimes 
(I_4 \otimes I_6))U_{(1)23\hat{4}\hat{5}(6)}
(\tilde{\rho}_{12} \otimes \chi_{34} \otimes \chi_{56})\right.
$$
$$
\left.
U^{\dagger}_{(1)23\hat{4}\hat{5}(6)}
((\tilde{\Pi}^{ij}_{13} \otimes \tilde{\Pi}^{i'j'}_{25}) \otimes 
(I_4 \otimes I_6))^{\dagger}
\right]
$$
Here, the unitary swap operator $U_{(1)23\hat{4}\hat{5}(6)}$ acts on product states according to
$$
U_{(1)23\hat{4}\hat{5}(6)}(
|\tilde{a}\rangle_1 \otimes |\tilde{b}\rangle_2 \otimes |c\rangle_3 \otimes |d\rangle_4 \otimes |e\rangle_5 \otimes |f\rangle_6) =
|\tilde{a}\rangle_1 \otimes |\tilde{c}\rangle_2 \otimes |b\rangle_3 \otimes |e\rangle_4 \otimes |d\rangle_5 \otimes |f\rangle_6
$$
swapping the states of systems $2$ and $3$, and those of systems $4$ and $5$, while leaving systems $1$ and $6$ alone.  It serves to cast the expression for $\rho^{ii'}_{46}$ into a form similar to that of Eq.(3).  The other unitary swap operator $U_{1(2)\hat{3}(4)5\hat{6}}$, acts on product states according to
$$
U_{1(2)\hat{3}(4)5\hat{6}}(
|\tilde{a}\rangle_1 \otimes |\tilde{c}\rangle_2 \otimes |b\rangle_3 \otimes |e\rangle_4 \otimes |d\rangle_5 \otimes |f\rangle_6) =
|\tilde{d}\rangle_1 \otimes |\tilde{c}\rangle_2 \otimes |f\rangle_3 \otimes |e\rangle_4 \otimes |a\rangle_5 \otimes |b\rangle_6,
$$
swapping the states of systems $1$ and $5$, and those of systems $3$ and $6$, while leaving systems $2$ and $4$ alone.  It plays an analogous role to $U_{1(2)3}$ in Eq.(4), establishing a one-to-one correspondence from the state space of the composite system onto itself, since the systems $12$ and $46$ are identical and thus have the same state space.  Writing
$$
\tilde{\chi}_{13} = \sum_k q_k|\tilde{s}_k\rangle_{13}\langle\tilde{s}_k|,
$$
$$
\tilde{\chi}_{25} = \sum_{k'} q_{k'}|\tilde{s}_{k'}\rangle_{25}\langle\tilde{s}_{k'}|,
$$
where the vectors $|\tilde{s}_k\rangle_{13}$ $(|\tilde{s}_{k'}\rangle_{25})$ make up the complete orthonormal set of eigenvectors of $\tilde{\chi}_{13}$ $(\tilde{\chi}_{25})$ in the joint space of $1$ $(2)$ and $3$ $(5)$, and performing the partial trace of Eq.(7) in the complete orthonormal basis $|\tilde{P}_l\rangle_{13}$ $(|\tilde{P}_{l'}\rangle_{25})$ for the joint system $13$ $(25)$ give
$$
\rho^{ii'}_{46} = \frac{1}{p_{ii'}}\sum_{j, j', k, k', l, l'}q_kq_{k'}\ 
{_{13}}\langle\tilde{P}_l|{_{25}}\langle\tilde{P}_{l'}|
((\tilde{\Pi}^{ij}_{13} \otimes \tilde{\Pi}^{i'j'}_{25}) 
\otimes (I_4 \otimes I_6))
U_{1(2)\hat{3}(4)5\hat{6}}
$$
$$
(|\tilde{s}_k\rangle_{13}\langle\tilde{s}_k| \otimes 
 |\tilde{s}_{k'}\rangle_{25}\langle\tilde{s}_{k'}| \otimes \rho_{46})
U^{\dagger}_{1(2)\hat{3}(4)5\hat{6}}
((\tilde{\Pi}^{ij}_{13} \otimes \tilde{\Pi}^{i'j'}_{25}) 
\otimes (I_4 \otimes I_6))^{\dagger}
|\tilde{P}_l\rangle_{13}|\tilde{P}_{l'}\rangle_{25}
$$
$$
= \sum_{j, j', k, k', l, l'}\left[\sqrt{\frac{q_kq_{k'}}{p_{ii'}}}
{_{13}}\langle\tilde{P}_l|{_{25}}\langle\tilde{P}_{l'}|
((\tilde{\Pi}^{ij}_{13} \otimes \tilde{\Pi}^{i'j'}_{25}) \otimes (I_4 \otimes I_6))U_{1(2)\hat{3}(4)5\hat{6}}
|\tilde{s}_k\rangle_{13}|\tilde{s}_{k'}\rangle_{25}\right]\rho_{46}
$$
$$
\left[
\sqrt{\frac{q_kq_{k'}}{p_{ii'}}}
{_{13}}\langle\tilde{s}_k|{_{25}}\langle\tilde{s}_{k'}|
U^{\dagger}_{1(2)\hat{3}(4)5\hat{6}}
((\tilde{\Pi}^{ij}_{13} \otimes \tilde{\Pi}^{i'j'}_{25}) 
\otimes (I_4 \otimes I_6))^{\dagger}
|\tilde{P}_l\rangle_{13}|\tilde{P}_{l'}\rangle_{25}
\right].
$$
Therefore, $\rho^{ii'}_{46}$ is related to $\rho_{46}$ by a quantum operation ${\cal E}^{ii'}$:
\begin{equation}
\rho^{ii'}_{46} = {\cal E}^{ii'}(\rho_{46}) = \sum_m A^{ii'm}_{46}\rho_{46}A^{ii'm\dagger}_{46},
\end{equation}
where
$$
A^{ii'm}_{46} \equiv \sqrt{\frac{q_kq_{k'}}{p_{ii'}}}
{_{13}}\langle\tilde{P}_l|{_{25}}\langle\tilde{P}_{l'}|
((\tilde{\Pi}^{ij}_{13} \otimes \tilde{\Pi}^{i'j'}_{25}) 
\otimes (I_4 \otimes I_6))U_{1(2)\hat{3}(4)5\hat{6}}
|\tilde{s}_k\rangle_{13}|\tilde{s}_{k'}\rangle_{25},
$$
\begin{equation}
\sum_m A^{ii'm\dagger}_{46}A^{ii'm}_{46} \leq I_{46},
\end{equation}
and the single index $m$ denotes the sextuple $(j, j', k, k', l, l')$.

Now, we are ready to show that, for two-level systems $1$, $2$, $3$, $4$, $5$ and $6$ (from hereon we consider only two-level systems), with
$$
\chi_{34} = 
q_1|\Phi^+\rangle_{34}\langle\Phi^+| +
q_2|\Phi^-\rangle_{34}\langle\Phi^-| +
q_3|\Psi^+\rangle_{34}\langle\Psi^+| +
q_4|\Psi^-\rangle_{34}\langle\Psi^-|,
$$
\begin{equation}
\chi_{56} = 
q_1|\Phi^+\rangle_{56}\langle\Phi^+| +
q_2|\Phi^-\rangle_{56}\langle\Phi^-| +
q_3|\Psi^+\rangle_{56}\langle\Psi^+| +
q_4|\Psi^-\rangle_{56}\langle\Psi^-|
\end{equation}
where $0 \leq q_k \leq 1$, $\sum^4_{k = 1}q_k = 1$,
$$
|\Phi^{\pm}\rangle = \frac{1}{\sqrt{2}}(|00\rangle \pm |11\rangle),
$$
$$
|\Psi^{\pm}\rangle = \frac{1}{\sqrt{2}}(|01\rangle \pm |10\rangle),
$$
are the Bell states, and
$$
\tilde{\Pi}^{1j}_{13} = \tilde{\Pi}^1_{13} =
\tilde{\Pi}^{1j'}_{25} = \tilde{\Pi}^1_{25} =
|\tilde{\Phi}^+\rangle\langle\tilde{\Phi}^+|,\
\tilde{\Pi}^{2j}_{13} = \tilde{\Pi}^2_{13} =
\tilde{\Pi}^{2j'}_{25} = \tilde{\Pi}^2_{25} =
|\tilde{\Phi}^-\rangle\langle\tilde{\Phi}^-|,
$$
\begin{equation}
\tilde{\Pi}^{3j}_{13} = \tilde{\Pi}^3_{13} = 
\tilde{\Pi}^{3j'}_{25} = \tilde{\Pi}^3_{25} =
|\tilde{\Psi}^+\rangle\langle\tilde{\Psi}^+|,\
\tilde{\Pi}^{4j}_{13} = \tilde{\Pi}^4_{13} =
\tilde{\Pi}^{4j'}_{25} = \tilde{\Pi}^4_{25}
|\tilde{\Psi}^-\rangle\langle\tilde{\Psi}^-|,
\end{equation}
then ${\cal E}^{11}$ describes two uncorrelated generalized depolarizing channels.  Here, we use $|0\rangle$ and $|1\rangle$ to denote an orthonormal set of basis states for each two-level system.  Eq.(10) and Eq.(11) allow us to calculate $p_{ii'} = \frac{1}{16}$ for all $1 \leq i, i' \leq 4$.  Substituting Eq.(10) into Eq.(7), we obtain
$$
\tilde{\chi}_{13} = 
q_1|\tilde{\Phi}^+\rangle_{13}\langle\tilde{\Phi}^+| +
q_2|\tilde{\Phi}^-\rangle_{13}\langle\tilde{\Phi}^-| +
q_3|\tilde{\Psi}^+\rangle_{13}\langle\tilde{\Psi}^+| +
q_4|\tilde{\Psi}^-\rangle_{13}\langle\tilde{\Psi}^-|,
$$
$$
\tilde{\chi}_{25} = 
q_1|\tilde{\Phi}^+\rangle_{25}\langle\tilde{\Phi}^+| +
q_2|\tilde{\Phi}^-\rangle_{25}\langle\tilde{\Phi}^-| +
q_3|\tilde{\Psi}^+\rangle_{25}\langle\tilde{\Psi}^+| +
q_4|\tilde{\Psi}^-\rangle_{25}\langle\tilde{\Psi}^-|.
$$
That is, the complete orthonormal set of eigenvectors of $\tilde{\chi}_{13}$ $(\tilde{\chi}_{25})$ in the joint space of $1$ $(2)$ and $3$ $(5)$ is given by
$$
|\tilde{s}_1\rangle_{13} = |\tilde{\Phi}^+\rangle_{13},\
|\tilde{s}_2\rangle_{13} = |\tilde{\Phi}^-\rangle_{13},\
|\tilde{s}_3\rangle_{13} = |\tilde{\Psi}^+\rangle_{13},\
|\tilde{s}_4\rangle_{13} = |\tilde{\Psi}^-\rangle_{13},
$$
\begin{equation}
|\tilde{s}_1\rangle_{25} = |\tilde{\Phi}^+\rangle_{25},\
|\tilde{s}_2\rangle_{25} = |\tilde{\Phi}^-\rangle_{25},\
|\tilde{s}_3\rangle_{25} = |\tilde{\Psi}^+\rangle_{25},\
|\tilde{s}_4\rangle_{25} = |\tilde{\Psi}^-\rangle_{25}.
\end{equation}
Eq.(9), when $i = i' = 1$, is in this case reduced to
\begin{equation}
A^{11kk'll'}_{46} = \sqrt{\frac{q_kq_{k'}}{p_{11}}}
{_{13}}\langle\tilde{P}_l|{_{25}}\langle\tilde{P}_{l'}|
((\tilde{\Pi}^1_{13} \otimes \tilde{\Pi}^1_{25}) \otimes (I_4 \otimes I_6))
U_{1(2)\hat{3}(4)5\hat{6}}
|\tilde{s}_k\rangle_{13}|\tilde{s}_{k'}\rangle_{25}.
\end{equation}
Making the choice
$$
|\tilde{P}_1\rangle_{13} = |\tilde{\Phi}^+\rangle_{13},\
|\tilde{P}_2\rangle_{13} = |\tilde{\Phi}^-\rangle_{13},\
|\tilde{P}_3\rangle_{13} = |\tilde{\Psi}^+\rangle_{13},\
|\tilde{P}_4\rangle_{13} = |\tilde{\Psi}^-\rangle_{13},
$$
\begin{equation}
|\tilde{P}_1\rangle_{25} = |\tilde{\Phi}^+\rangle_{25},\
|\tilde{P}_2\rangle_{25} = |\tilde{\Phi}^-\rangle_{25},\
|\tilde{P}_3\rangle_{25} = |\tilde{\Psi}^+\rangle_{25},\
|\tilde{P}_4\rangle_{25} = |\tilde{\Psi}^-\rangle_{25},
\end{equation}
and substituting Eq.(12) and Eq.(14) into Eq.(13) yield two uncorrelated generalized depolarizing channels ${\cal E}^{11}$ specified by Kraus operators
$$
A^{111111}_{46} = \sqrt{q_1}\sqrt{q_1}I_4 \otimes I_6,\
A^{111211}_{46} = \sqrt{q_1}\sqrt{q_2}I_4 \otimes \sigma^z_6,
$$
$$
A^{111311}_{46} = \sqrt{q_1}\sqrt{q_3}I_4 \otimes \sigma^x_6,\
A^{111411}_{46} = \sqrt{q_1}\sqrt{q_4}I_4 \otimes \sigma^y_6,
$$
$$
A^{112111}_{46} = \sqrt{q_2}\sqrt{q_1}\sigma^z_4 \otimes I_6,\
A^{112211}_{46} = \sqrt{q_2}\sqrt{q_2}\sigma^z_4 \otimes \sigma^z_6,
$$
$$
A^{112311}_{46} = \sqrt{q_2}\sqrt{q_3}\sigma^z_4 \otimes \sigma^x_6,\
A^{112411}_{46} = \sqrt{q_2}\sqrt{q_4}\sigma^z_4 \otimes \sigma^y_6,
$$
$$
A^{113111}_{46} = \sqrt{q_3}\sqrt{q_1}\sigma^x_4 \otimes I_6,\
A^{113211}_{46} = \sqrt{q_3}\sqrt{q_2}\sigma^x_4 \otimes \sigma^z_6,
$$
$$
A^{113311}_{46} = \sqrt{q_3}\sqrt{q_3}\sigma^x_4 \otimes \sigma^x_6,\
A^{113411}_{46} = \sqrt{q_3}\sqrt{q_4}\sigma^x_4 \otimes \sigma^y_6,
$$
$$
A^{114111}_{46} = \sqrt{q_4}\sqrt{q_1}\sigma^y_4 \otimes I_6,\
A^{114211}_{46} = \sqrt{q_4}\sqrt{q_2}\sigma^y_4 \otimes \sigma^z_6,
$$
\begin{equation}
A^{114311}_{46} = \sqrt{q_4}\sqrt{q_3}\sigma^y_4 \otimes \sigma^x_6,\
A^{114411}_{46} = \sqrt{q_4}\sqrt{q_4}\sigma^y_4 \otimes \sigma^y_6,
\end{equation}
where
$$
\sigma^x = \left(\begin{array}{cc} 0 & 1 \\ 1 & 0 \end{array}\right),\
\sigma^y = \left(\begin{array}{cc} 0 & -i\\ i & 0 \end{array}\right),\
\sigma^z = \left(\begin{array}{cc} 1 & 0 \\ 0 & -1\end{array}\right)
$$
are the Pauli matrices.  It turns out that, with different Alice's measurement outcomes $(i, i')$ we end up with different two uncorrelated generalized depolarizing channels ${\cal E}^{ii'}$.  These channels differ only in the coefficients $\sqrt{q_k}\sqrt{q_{k'}}$ in front of $I_4 \otimes I_6$, $I_4 \otimes \sigma^z_6$, etc.

For Bob to successfully complete the teleportation protocol, he must perform a $(i, i')$-dependent trace-preserving quantum operation ${\cal R}^{ii'}$:
$$
{\cal R}^{ii'}(\rho^{ii'}_{46}) = 
\sum_n B^{ii'n}_{46}\rho^{ii'}_{46}B^{ii'n\dagger}_{46},
$$
\begin{equation}
\sum_n B^{ii'n\dagger}_{46}B^{ii'n}_{46} = I_{46},
\end{equation}
such that the fidelity \cite{Jozsa}
$
F^{ii'}(\rho_{46}, {\cal R}^{ii'}\circ{\cal E}^{ii'}(\rho_{46}))
$
between the input state $\rho_{46}$ and the teleported state ${\cal R}^{ii'}\circ{\cal E}^{ii'}(\rho_{46})$ is optimal, that is, as close to one as possible.  In other words, Bob has to determine ${\cal R}^{ii'}$ which optimally reverses ${\cal E}^{ii'}$.  For simplicity, we assume the input state $\rho_{46}$ is an entangled pure state:
\begin{equation}
\rho_{46} = (\cos\theta|00\rangle + \sin\theta|11\rangle)_{46}
(\langle 00|\cos\theta + \langle 11|\sin\theta),
\end{equation}
with $E(\rho_{46}) = \sin 2\theta$.  In Ref.\cite{Yeo}, it was shown that the optimal approximate reversal of a generalized depolarizing channel can be achieved using only unitary transformations: $I, \sigma^x, \sigma^y, \sigma^z$.  For instance, since ${\cal E}^{11}$ describes two uncorrelated generalized depolarizing channels, we have
$$
{\cal R}^{11}(\rho^{11}_{46}) = 
B^{11}_{46}\rho^{11}_{46}B^{11\dagger}_{46},
$$
where $B^{11}_{46}$ is composed of a tensor product of two unitary operators: $I, \sigma^x, \sigma^y, \sigma^z$.  The exact expression for $B^{11}_{46}$ is determined by the relative magnitudes of $q_1, q_2, q_3$ and $q_4$.  After some algebra, we obtain the maximum fidelity
$$
F_{\max} = \sum^4_{i, i' = 1}p_{ii'}F^{ii'}_{\max}(\rho_{46}, {\cal R}^{ii'}\circ{\cal E}^{ii'}(\rho_{46})) = \sum^4_{i, i' = 1}p_{ii'}
{\rm tr}_{46}(\rho_{46}, {\cal R}^{ii'}\circ{\cal E}^{ii'}(\rho_{46}))
$$
$$
= \max((q_1 + q_2)^2 + (q_1^2 - 2q_3q_4 + q_2^2)\sin^22\theta,
$$
$$
(q_1 + q_2)^2 + (q_3^2 - 2q_1q_2 + q_4^2)\sin^22\theta,
$$
$$
(q_3 + q_4)^2 + (q_3^2 - 2q_1q_2 + q_4^2)\sin^22\theta,
$$
$$
(q_3 + q_4)^2 + (q_1^2 - 2q_3q_4 + q_2^2)\sin^22\theta,
$$
$$
(q_1 + q_2)(q_3 + q_4) + (q_1 - q_2)(q_3 - q_4)\sin^22\theta,
$$
\begin{equation}
(q_1 + q_2)(q_3 + q_4) - (q_1 - q_2)(q_3 - q_4)\sin^22\theta).
\end{equation}
We note that for $q_1 = q_2 = q_3 = \frac{1 - \phi}{6}$, and $q_4 = \frac{1 + \phi}{2}$, Eq.(18) yields \cite{Lee}
\begin{equation}
F_{\max} = \frac{1}{9}[(2 + \phi)^2 - (1 - \phi)(1 + 2\phi)\sin^22\theta]
\end{equation}
Also, the amount of entanglement associated with the teleported state is given by
\begin{equation}
E({\cal R}^{ii'}\circ{\cal E}^{ii'}(\rho_{46})) = \max(0, \frac{1}{9}[(1 + 2\phi)^2\sin 2\theta]- 2(2 - \phi - \phi^2)).
\end{equation}

It is impossible to obtain two correlated generalized depolarizing channels from Eq.(7).  This is because, by insisting on $\chi_{34} \otimes \chi_{56}$, and demanding that the resulting Kraus operators reduces to Eq.(25), would in general result in requiring Alice's generalized measurements not be of the physically meaningful form in Eq.(7).  So, we keep Eq.(11), and instead of Eq.(10), consider
$$
\chi_{3456} = |\chi\rangle_{3456}\langle\chi|,
$$
\begin{equation}
|\chi\rangle_{3456} \equiv
\sqrt{q_1}|\Phi^+\rangle_{34} \otimes |\Phi^+\rangle_{56} +
\sqrt{q_2}|\Phi^-\rangle_{34} \otimes |\Phi^-\rangle_{56} +
\sqrt{q_3}|\Psi^+\rangle_{34} \otimes |\Psi^+\rangle_{56} +
\sqrt{q_4}|\Psi^-\rangle_{34} \otimes |\Psi^-\rangle_{56},
\end{equation}
where $0 \leq q_k \leq 1$, $\sum^4_{k = 1} q_k = 1$.  Replacing $\chi_{34} \otimes \chi_{56}$ in Eq.(7) by $\chi_{3456}$, we obtain
$$
\tilde{\chi}_{1325} = |\tilde{\chi}\rangle_{1325}\langle\tilde{\chi}|,
$$
\begin{equation}
|\tilde{\chi}\rangle_{1325} \equiv
\sqrt{q_1}|\tilde{\Phi}^+\rangle_{13} \otimes |\tilde{\Phi}^+\rangle_{25} +
\sqrt{q_2}|\tilde{\Phi}^-\rangle_{13} \otimes |\tilde{\Phi}^-\rangle_{25} +
\sqrt{q_3}|\tilde{\Psi}^+\rangle_{13} \otimes |\tilde{\Psi}^+\rangle_{25} +
\sqrt{q_4}|\tilde{\Psi}^-\rangle_{13} \otimes |\tilde{\Psi}^-\rangle_{25}.
\end{equation}
Consequently, when $i = i' = 1$, we obtain, instead of Eq.(13),
\begin{equation}
A^{11kll'}_{46} = \sqrt{\frac{q_k}{p_{11}}}
{_{13}}\langle\tilde{P}_l|{_{25}}\langle\tilde{P}_{l'}|
((\tilde{\Pi}^1_{13} \otimes \tilde{\Pi}^1_{25}) \otimes (I_4 \otimes I_6))
U_{1(2)\hat{3}(4)5\hat{6}}
|\tilde{s}_k\rangle_{13}|\tilde{s}_k\rangle_{25}
\end{equation}
where
\begin{equation}
|\tilde{s}_1\rangle = |\tilde{\Phi}^+\rangle,\
|\tilde{s}_2\rangle = |\tilde{\Phi}^-\rangle,\
|\tilde{s}_3\rangle = |\tilde{\Psi}^+\rangle,\
|\tilde{s}_4\rangle = |\tilde{\Psi}^-\rangle.
\end{equation}
Substituting Eq.(14) and Eq.(24) into Eq.(23) gives two correlated generalized depolarizing channels ${\cal E}^{11}$ specified by Kraus operators
$$
A^{11111}_{46} = \sqrt{\frac{q_1}{\pi_{11}}}I_4 \otimes I_6,\
A^{11211}_{46} = \sqrt{\frac{q_2}{\pi_{11}}}\sigma^z_4 \otimes \sigma^z_6,
$$
\begin{equation}
A^{11311}_{46} = \sqrt{\frac{q_3}{\pi_{11}}}\sigma^x_4 \otimes \sigma^x_6,\
A^{11411}_{46} = \sqrt{\frac{q_4}{\pi_{11}}}\sigma^y_4 \otimes \sigma^y_6.
\end{equation}
with probability $p_{11} = \frac{1}{16}\pi_{11}$, where
\begin{equation}
\pi_{11} = 1 + 
2(\sqrt{q_1}\sqrt{q_3} + \sqrt{q_2}\sqrt{q_4})c_{11} -
2(\sqrt{q_1}\sqrt{q_4} + \sqrt{q_2}\sqrt{q_3})c_{22} +
2(\sqrt{q_1}\sqrt{q_2} + \sqrt{q_3}\sqrt{q_4})c_{33}.
\end{equation}
Here, $c_{11}$, $c_{22}$ and $c_{33}$ are real coefficients in
$$
\rho_{46} = \frac{1}{4}(I_4 \otimes I_6 + \vec{a}\cdot\vec{\sigma}_4 \otimes I_6 + \vec{b}\cdot I_4 \otimes \vec{\sigma}_6 + \sum^3_{r, s = 1}c_{rs}\sigma^r_4 \otimes \sigma^s_6)
$$
with $\sigma^x \equiv \sigma^1$, $\sigma^y \equiv \sigma^2$, and $\sigma^z \equiv \sigma^3$.  Therefore, information about the input state could be obtained, in accordance with Ref.\cite{Bana}.  However, we note that it is only information about the associated entanglement encoded via $c_{11}$, $c_{22}$, and $c_{33}$ which is obtainable.  Information about the individual subsystems $4$ and $6$, and that about the associated entanglement encoded via $c_{rs},\ r\not= s$, is not.

It turns out that, for different $i = i'$, we have different two correlated generalized depolarizing channels ${\cal E}^{ii}$ with probability $p_{ii} = p_{11}$.  Again, these channels differ only in the coefficients $\sqrt{q_k}$.  For $i \not= i'$, we do not have two correlated generalized depolarizing channels specified by Kraus operators of the form in Eq.(25).  However, there exsists ``partial correlations'', and the Kraus operators for these channels can be evaluated similarly.  The corresponding probabilities are
$$
p_{12} = p_{21} = p_{34} = p_{43} = \frac{1}{16}\pi_{12},
$$
$$
\pi_{12} = 1 - 
2(\sqrt{q_1}\sqrt{q_3} + \sqrt{q_2}\sqrt{q_4})c_{11} +
2(\sqrt{q_1}\sqrt{q_4} + \sqrt{q_2}\sqrt{q_3})c_{22} +
2(\sqrt{q_1}\sqrt{q_2} + \sqrt{q_3}\sqrt{q_4})c_{33},
$$
$$
p_{13} = p_{24} = p_{31} = p_{42} = \frac{1}{16}\pi_{12},
$$
$$
\pi_{13} = 1 + 
2(\sqrt{q_1}\sqrt{q_3} + \sqrt{q_2}\sqrt{q_4})c_{11} +
2(\sqrt{q_1}\sqrt{q_4} + \sqrt{q_2}\sqrt{q_3})c_{22} -
2(\sqrt{q_1}\sqrt{q_2} + \sqrt{q_3}\sqrt{q_4})c_{33},
$$
$$
p_{14} = p_{23} = p_{32} = p_{41} = \frac{1}{16}\pi_{12},
$$
\begin{equation}
\pi_{14} = 1 - 
2(\sqrt{q_1}\sqrt{q_3} + \sqrt{q_2}\sqrt{q_4})c_{11} -
2(\sqrt{q_1}\sqrt{q_4} + \sqrt{q_2}\sqrt{q_3})c_{22} -
2(\sqrt{q_1}\sqrt{q_2} + \sqrt{q_3}\sqrt{q_4})c_{33}.
\end{equation}
Assuming the input state is given by Eq.(17), and employing the above unitary ${\cal R}^{ii'}$, we obtain, after some algebra, the maximum fidelity
\begin{equation}
F_{\max} = \max(
(q_1 + q_2) + (q_3 + q_4)\sin^22\theta, (q_3 + q_4) + (q_1 + q_2)\sin^22\theta).
\end{equation}
This is in contrast to Eq.(19) since $F_{\max}$ increases with the amount of entanglement associated with the input state.  More interestingly, when we calculate the amount of entanglement associated with the teleported state, we find
\begin{equation}
E({\cal R}^{ii'}\circ{\cal E}^{ii'}(\rho_{46})) = \sin 2\theta,
\end{equation}
equal to the amount of entanglement associated with the input state, Eq.(17).

In conclusion, we have demonstrated how the entanglement teleportation scheme \cite{Lee} gives rise to two uncorrelated generalized depolarizing channels.  In an attempt to find a teleportation scheme which yields two correlated generalized depolarizing channels, we discover a novel teleportation scheme which allows us to learn about the entanglement in the entangled pure input state, but without decreasing the amount of entanglement associated with it.

The author thanks Yuri Suhov and Andrew Skeen for useful discussions.  This publication is an output from project activity funded by The Cambridge MIT Institute Limited (``CMI'').  CMI is funded in part by the United Kingdom Government.  The activity was carried out for CMI by the University of Cambridge and Massachusetts Institute of Technology.  CMI can accept no responsibility for any information provided or views expressed.

\end{document}